\input harvmac.tex
\noblackbox
\font\ticp=cmcsc10

\font\cmss=cmss10


\def\unlockat{\catcode`\@=11}
\def\lockat{\catcode`\@=12}

\unlockat

\def\newsec#1{\global\advance\secno by1\message{(\the\secno. #1)}
\global\subsecno=0\global\subsubsecno=0\eqnres@t\noindent {\bf\the\secno.
#1}
\writetoca{{\secsym} {#1}}\par\nobreak\medskip\nobreak}
\global\newcount\subsecno \global\subsecno=0
\def\subsec#1{\global\advance\subsecno by1\message{(\secsym\the\subsecno.
#1)}
\ifnum\lastpenalty>9000\else\bigbreak\fi\global\subsubsecno=0
\noindent{\it\secsym\the\subsecno. #1}
\writetoca{\string\quad {\secsym\the\subsecno.} {#1}}
\par\nobreak\medskip\nobreak}
\global\newcount\subsubsecno \global\subsubsecno=0
\def\subsubsec#1{\global\advance\subsubsecno by1
\message{(\secsym\the\subsecno.\the\subsubsecno. #1)}
\ifnum\lastpenalty>9000\else\bigbreak\fi
\noindent\quad{\secsym\the\subsecno.\the\subsubsecno.}{#1}
\writetoca{\string\qquad{\secsym\the\subsecno.\the\subsubsecno.}{#1}}
\par\nobreak\medskip\nobreak}

\def\subsubseclab#1{\DefWarn#1\xdef
#1{\noexpand\hyperref{}{subsubsection}%
{\secsym\the\subsecno.\the\subsubsecno}%
{\secsym\the\subsecno.\the\subsubsecno}}%
\writedef{#1\leftbracket#1}\wrlabeL{#1=#1}}
\lockat

%
%
%

\def\CH{{\cal H}}

\def\CM{{\cal M}}
\def\cM{{\cal M}}

\def\CV{{\cal V}}

\def\h{{\cal H}}

\def\IZ{\relax\ifmmode\mathchoice {\hbox{\cmss Z\kern-.4em
Z}}{\hbox{\cmss Z\kern-.4em Z}} {\lower.9pt\hbox{\cmsss Z\kern-.4em Z}}
{\lower1.2pt\hbox{\cmsss Z\kern-.4em Z}}\else{\cmss Z\kern-.4em Z}\fi}
\def\IB{\relax{\rm I\kern-.18em B}}
\def\IC{{\relax\hbox{$\inbar\kern-.3em{\rm C}$}}}
\def\ID{\relax{\rm I\kern-.18em D}}
\def\IE{\relax{\rm I\kern-.18em E}}
\def\IF{\relax{\rm I\kern-.18em F}}
\def\IG{\relax\hbox{$\inbar\kern-.3em{\rm G}$}}
\def\IGa{\relax\hbox{${\rm I}\kern-.18em\Gamma$}}
\def\IH{\relax{\rm I\kern-.18em H}}
\def\II{\relax{\rm I\kern-.18em I}}
\def\IK{\relax{\rm I\kern-.18em K}}
\def\IP{\relax{\rm I\kern-.18em P}}

\def\inbar{\,\vrule height1.5ex width.4pt depth0pt}

\font\cmss=cmss10 \font\cmsss=cmss10 at 7pt
\def\IR{\relax{\rm I\kern-.18em R}}


\def\bS{{\bf S}}
\def\bT{{\bf T}}
\def\bX{{\bf X}}


\def\II{{\unit}}

\def\rsemidir{\mathbin{\hbox{$\times$\hskip-2pt\vrule height 5.7pt depth -.3pt
width .25pt\hskip2pt}}}


 \def\unit{\hbox to 3.3pt{\hskip1.3pt \vrule height 7pt width .4pt
\hskip.7pt
\vrule height 7.85pt width .4pt \kern-2.4pt
\hrulefill \kern-3pt
\raise 4pt\hbox{\char'40}}}

\def\t{{\tau}}

\def\tone{{\tau_1}}
\def\ttwo{{\tau_2}}

\font\mybb=msbm10 at 10pt
\def\bbbb#1{\hbox{\mybb#1}}
\def\Z {\bbbb{Z}}
\def\R {\bbbb{R}}
\def\bZ{{\bf Z}}
\def\bK{{\bf K}}

%

\def\rsemidir{\mathbin{\hbox{$\times$\hskip-2pt\vrule height 5.7pt depth -.3pt 
width
.25pt\hskip2pt}}}


\def \ll {\lambda}

\def \ss {\sigma}

\def \lll {\Lambda}

\def \ti {\tilde}

\def \2 {{1 \over 2}}
\def \3 {{1 \over 3}}
\def \4 {{1 \over 4}}
\def \5 {{1 \over 5}}
\def \6 {{1 \over 6}}
\def \7 {{1 \over 7}}
\def \8 {{1 \over 8}}
\def \9 {{1 \over 9}}
\def \0 { \infty}

\def\++ {{(+)}}
\def \- {{(-)}}
\def\+-{{(\pm)}}

\def \qq {\qquad}

\def\npb#1#2#3{{\sl Nucl. Phys.} {\bf B#1} (#2) #3}

\lref\deBoerPX{
J.~de Boer, R.~Dijkgraaf, K.~Hori, A.~Keurentjes, J.~Morgan, 
D.~R.~Morrison and S.~Sethi,
``Triples, fluxes, and strings,''
Adv.\ Theor.\ Math.\ Phys.\  {\bf 4}, 995 (2002)
[arXiv:hep-th/0103170].
}

\lref\PolchinskiSM{
J.~Polchinski and A.~Strominger,
``New Vacua for Type II String Theory,''
Phys.\ Lett.\ B {\bf 388}, 736 (1996)
[arXiv:hep-th/9510227].
}

\lref\MichelsonPN{
J.~Michelson,
``Compactifications of type IIB strings to four dimensions 
with  non-trivial classical potential,''
Nucl.\ Phys.\ B {\bf 495}, 127 (1997)
[arXiv:hep-th/9610151].
}

\lref\CurioSC{
G.~Curio, A.~Klemm, D.~Lust and S.~Theisen,
``On the vacuum structure of type II string compactifications on  Calabi-Yau spaces with H-fluxes,''
Nucl.\ Phys.\ B {\bf 609}, 3 (2001)
[arXiv:hep-th/0012213].
}

\lref\GreeneGH{
B.~R.~Greene, K.~Schalm and G.~Shiu,
``Warped compactifications in M and F theory,''
Nucl.\ Phys.\ B {\bf 584}, 480 (2000)
[arXiv:hep-th/0004103].
}

\lref\DasguptaSS{
K.~Dasgupta, G.~Rajesh and S.~Sethi,
``M theory, orientifolds and G-flux,''
JHEP {\bf 9908}, 023 (1999)
[arXiv:hep-th/9908088].
}

\lref\Bank{For a review see T. Banks, ``SUSY Breaking, Cosmology, Vacuum
Selection and the Cosmological Constant in String Theory,''
hep-th/9601151.}

\lref\MuWi{M. Mueller and E. Witten, ``Twisting Toroidally Compactified
Heterotic Strings with Enlarged Symmetry Groups,'' {\sl Phys. Lett.} {\bf
182B} (1986) 28.}

\lref\HMV{J. A. Harvey. G. Moore and C. Vafa, ``Quasicrystalline
Compactification,''  \npb{304}{1988}{269}.}

\lref\Nara{K. S. Narain, ``New Heterotic String Theories in Uncompactified
Dimension $< 10$,'' { \sl Phys. Lett.} {\bf 169B} (1986) 41.}

\lref\DiSi{M. Dine and E. Silverstein, ``New M-theory Backgrounds with
Frozen Moduli,'' hep-th/9712166.}

\lref \ScSc{J. Scherk and J.H. Schwarz, ``How to Get Masses from Extra 
Dimensions,'' {Nucl. Phys. B}~{\bf 153} (1979) 61.}

\lref\Berg{E. Bergshoeff, M. de Roo, M.B. Green, G. Papadopoulos and P.K. 
Townsend, ``Duality of Type-II 7-branes and 8-branes,''
{Nucl. Phys. B} {\bf 470}, 113 (1996), [arXiv:hep-th/9601150].}

\lref\fibs{ I.V.~Lavrinenko, H.~Lu and C.N.~Pope,  ``Fibre Bundles
and Generalised Dimensional Reductions,'' Class.\ Quant.\
Grav.\  {\bf 15}, 2239 (1998), [arXiv:hep-th/9710243].}

\lref\kuri{
N.~Kaloper, R.~R.~Khuri and R.~C.~Myers,
``On generalized axion reductions,''
Phys.\ Lett.\ B {\bf 428}, 297 (1998)
[arXiv:hep-th/9803066].
}

\lref\gage{
E.~Bergshoeff, M.~de Roo and E.~Eyras,
``Gauged supergravity from dimensional reduction,''
Phys.\ Lett.\ B {\bf 413}, 70 (1997)
[arXiv:hep-th/9707130].
}

\lref\wall{
P.~M.~Cowdall, H.~Lu, C.~N.~Pope, K.~S.~Stelle and P.~K.~Townsend,
``Domain walls in massive supergravities,''
Nucl.\ Phys.\ B {\bf 486}, 49 (1997)
[arXiv:hep-th/9608173].
}

\lref\KaMy{ N. Kaloper and R. Myers, ``The O(dd) story of massive
supergravity,'' JHEP {\bf 05} (1999) 010,  [arXiv:hep-th{9901045}]. }

\lref\mf{C.M. Hull,  ``Massive string theories from M-theory and
F-theory,''   {JHEP} {\bf 9811} 027 (1998), [arXiv:hep-th/9811021].}

\lref\cow{
P.~M.~Cowdall, ``Novel domain wall and Minkowski vacua of D = 9
maximal SO(2) gauged supergravity,'' 
arXiv:hep-th/0009016.
}

\lref\ort{
P.~Meessen and T.~Ortin,
``An SL(2,Z) multiplet of nine-dimensional type II supergravity theories,''
Nucl.\ Phys.\ B {\bf 541}, 195 (1999)
[arXiv:hep-th/9806120].
}

\lref\HaackIZ{
M.~Haack, J.~Louis and H.~Singh,
``Massive type IIA theory on K3,''
JHEP {\bf 0104}, 040 (2001)
[arXiv:hep-th/0102110].
}

\lref\FerraraJX{
S.~Ferrara, C.~Kounnas, M.~Porrati and F.~Zwirner,
``Superstrings With Spontaneously Broken Supersymmetry And Their 
Effective Theories,''
Nucl.\ Phys.\ B {\bf 318}, 75 (1989).
}

\lref\FerraraES{
S.~Ferrara, C.~Kounnas and M.~Porrati,
``Superstring Solutions With Spontaneously Broken Four-Dimensional 
Supersymmetry,''
Nucl.\ Phys.\ B {\bf 304}, 500 (1988).
}

\lref\RohmAQ{
R.~Rohm,
``Spontaneous Supersymmetry Breaking In Supersymmetric String Theories,''
Nucl.\ Phys.\ B {\bf 237}, 553 (1984).
}

\lref\mee{
J.~Gheerardyn and P.~Meessen,
``Supersymmetry of massive D = 9 supergravity,''
Phys.\ Lett.\ B {\bf 525}, 322 (2002),
[arXiv:hep-th/0111130].}

\lref\HullWG{ C.~M.~Hull,
``Gauged D = 9 supergravities and Scherk-Schwarz reduction,''
arXiv:hep-th/0203146.}

\lref\bergnew{
E.~Bergshoeff, U.~Gran and D.~Roest,
``Type IIB seven-brane solutions from nine-dimensional domain walls,''
arXiv:hep-th/0203202.
}

\lref\DeWolfePR{
O.~DeWolfe, T.~Hauer, A.~Iqbal and B.~Zwiebach,
``Uncovering infinite symmetries on (p,q) 7-branes: 
Kac-Moody algebras  and beyond,''
Adv.\ Theor.\ Math.\ Phys.\  {\bf 3}, 1835 (1999)
[arXiv:hep-th/9812209].
}

\lref\DeWolfeEU{
O.~DeWolfe, T.~Hauer, A.~Iqbal and B.~Zwiebach,
``Uncovering the symmetries on (p,q) 7-branes: 
Beyond the Kodaira  classification,''
Adv.\ Theor.\ Math.\ Phys.\  {\bf 3}, 1785 (1999)
[arXiv:hep-th/9812028].
}

\lref\DabholkarKV{
A.~Dabholkar and J.~A.~Harvey,
``String islands,''
JHEP {\bf 9902}, 006 (1999)
[arXiv:hep-th/9809122].
}

\lref\GiveonFU{
A.~Giveon, M.~Porrati and E.~Rabinovici,
``Target space duality in string theory,''
Phys.\ Rept.\  {\bf 244}, 77 (1994)
[arXiv:hep-th/9401139].
}

\lref\BeckerSX{
K.~Becker and K.~Dasgupta,
arXiv:hep-th/0209077.
}

\lref\Oz{C.M. Hull and A.C. Ozer, in preparation.}
\lref\NarainAM{
K.~S.~Narain, M.~H.~Sarmadi and E.~Witten,
``A Note On Toroidal Compactification Of Heterotic String Theory,''
Nucl.\ Phys.\ B {\bf 279}, 369 (1987).
}

\lref\VafaGM{
C.~Vafa and E.~Witten,
``Dual string pairs with N = 1 and N = 2 supersymmetry in four  dimensions,''
Nucl.\ Phys.\ Proc.\ Suppl.\  {\bf 46}, 225 (1996)
[arXiv:hep-th/9507050].
}

\lref\DixonJW{
L.~J.~Dixon, J.~A.~Harvey, C.~Vafa and E.~Witten,
``Strings On Orbifolds,''
Nucl.\ Phys.\ B {\bf 261}, 678 (1985).
}

\lref\DabholkarZD{
A.~Dabholkar,
``Lectures on orientifolds and duality,''
arXiv:hep-th/9804208.
}

\lref\GanorZE{
O.~J.~Ganor,
``U-duality twists and possible phase transitions in (2+1)D  supergravity,''
Nucl.\ Phys.\ B {\bf 549}, 145 (1999)
[arXiv:hep-th/9812024].
}

\lref\Nara{K.~S.~Narain, Phys.\Lett.\ B {\bf 169}, 369 (1986).}

\lref\group{ S.~Helgason, ``Differential Geometry and Symmetric Spaces,''
Academic Press (1978), theorem (2.1), page 256.}

\lref\KumarZX{
A.~Kumar and C.~Vafa,
``U-manifolds,''
Phys.\ Lett.\ B {\bf 396}, 85 (1997)
[arXiv:hep-th/9611007].
}

\lref\HellermanAX{
S.~Hellerman, J.~McGreevy and B.~Williams,
``Geometric constructions of nongeometric string theories,''
arXiv:hep-th/0208174.
}

\lref\GiddingsYU{
S.~B.~Giddings, S.~Kachru and J.~Polchinski,
``Hierarchies from fluxes in string compactifications,''
arXiv:hep-th/0105097.
}

\lref\FreyHF{
A.~R.~Frey and J.~Polchinski,
``N = 3 warped compactifications,''
Phys.\ Rev.\ D {\bf 65}, 126009 (2002)
[arXiv:hep-th/0201029].
}

\lref\KachruHE{
S.~Kachru, M.~B.~Schulz and S.~Trivedi,
``Moduli stabilization from fluxes in a simple IIB orientifold,''
arXiv:hep-th/0201028.
}

\lref\GukovYA{
S.~Gukov, C.~Vafa and E.~Witten,
``CFT's from Calabi-Yau four-folds,''
Nucl.\ Phys.\ B {\bf 584}, 69 (2000)
[Erratum-ibid.\ B {\bf 608}, 477 (2001)]
[arXiv:hep-th/9906070].
}

\lref\LouisNY{
J.~Louis and A.~Micu,
``Type II theories compactified on Calabi-Yau threefolds 
in the presence  of background fluxes,''
Nucl.\ Phys.\ B {\bf 635}, 395 (2002)
[arXiv:hep-th/0202168].
}

\lref\LouisUY{
J.~Louis and A.~Micu,
``Heterotic string theory with background fluxes,''
Nucl.\ Phys.\ B {\bf 626}, 26 (2002)
[arXiv:hep-th/0110187].
}

\lref\HullYS{
C.~M.~Hull and P.~K.~Townsend,
``Unity of superstring dualities,''
Nucl.\ Phys.\ B {\bf 438}, 109 (1995)
[arXiv:hep-th/9410167].
}

\lref\CurioAE{
G.~Curio, A.~Klemm, B.~Kors and D.~Lust,
``Fluxes in heterotic and type II string compactifications,''
Nucl.\ Phys.\ B {\bf 620}, 237 (2002)
[arXiv:hep-th/0106155].
}

\lref\BeckerPM{
K.~Becker and M.~Becker,
``Supersymmetry breaking, M-theory and fluxes,''
JHEP {\bf 0107}, 038 (2001)
[arXiv:hep-th/0107044].
}

\lref\VafaXN{
C.~Vafa,
``Evidence for F-Theory,''
Nucl.\ Phys.\ B {\bf 469}, 403 (1996)
[arXiv:hep-th/9602022].
}

\lref\AndrianopoliAQ{
L.~Andrianopoli, R.~D'Auria, S.~Ferrara and M.~A.~Lledo,
``Duality and spontaneously broken supergravity in flat backgrounds,''
Nucl.\ Phys.\ B {\bf 640}, 63 (2002)
[arXiv:hep-th/0204145].
}

\lref\DAuriaTC{
R.~D'Auria, S.~Ferrara and S.~Vaula,
``N = 4 gauged supergravity and a IIB orientifold with fluxes,''
arXiv:hep-th/0206241.
}

\lref\RussoCV{
J.~G.~Russo and A.~A.~Tseytlin,
``Constant magnetic field in closed string theory: 
An Exactly solvable model,''
Nucl.\ Phys.\ B {\bf 448}, 293 (1995)
[arXiv:hep-th/9411099].
}

\lref\RussoIK{
J.~G.~Russo and A.~A.~Tseytlin,
``Magnetic flux tube models in superstring theory,''
Nucl.\ Phys.\ B {\bf 461}, 131 (1996)
[arXiv:hep-th/9508068].
}

\lref\TakayanagiJJ{
T.~Takayanagi and T.~Uesugi,
``Orbifolds as Melvin geometry,''
JHEP {\bf 0112}, 004 (2001)
[arXiv:hep-th/0110099].
}

%
%

\Title{\vbox{\baselineskip12pt
\hbox{hep-th/0210209}
\hbox{TIFR/TH/02-24}
\hbox{QMUL-PH-02-16} }} 
{\vbox{\centerline{Duality Twists, Orbifolds, and Fluxes}}}
\centerline{\ticp Atish Dabholkar $^{\dagger}$ 
and  Chris  Hull $^{\ddagger}$}
\bigskip
\centerline{$^{\dagger}$ \sl Department of Theoretical Physics}
\centerline{\sl Tata Institute of Fundamental Research}
\centerline{\sl Homi Bhabha Road, Mumbai, India 400005}
\centerline{Email: atish@tifr.res.in}
\bigskip
\centerline{$^{\ddagger}$ \sl Physics Department}
\centerline{\sl Queen Mary,  University of London}
\centerline{\sl Mile End Road, London,  E1 4NS, U. K.}
\centerline{Email: c.m.hull@qmul.ac.uk}
\bigskip
\centerline{ABSTRACT}
\medskip

We investigate compactifications with duality twists and their
relation to orbifolds and compactifications with fluxes. Inequivalent
compactifications are classified by conjugacy classes of the U-duality
group and result in gauged supergravities in lower dimensions with
nontrivial Scherk-Schwarz potentials on the moduli space. For certain
twists, this mechanism is equivalent to introducing internal fluxes
but is more general and can be used to stabilize some of the
moduli. We show that the potential has stable minima with zero energy
precisely at the fixed points of the twist group. In string theory,
when the twist belongs to the T-duality group, the theory at the
minimum has an exact CFT description as an orbifold.  We also discuss
more general twists by nonperturbative U-duality transformations.

\bigskip
\Date{October 2002}
%

\newsec{Introduction}

In this paper we investigate compactifications that include duality
twists and internal fluxes and their relation to orbifolds.

Compactification with duality twisting is a generalization of the
Scherk-Schwarz mechanism in classical supergravity \refs{\ScSc \Berg
\fibs \kuri \gage \wall \KaMy \mf \ort \cow \HullWG \HaackIZ {--} \bergnew}.  
In a typical supergravity theory, 
there is a non-compact global symmetry $G$.
In a twisted compactification, one introduces a twist in the
toroidal directions by the global symmetry $G$. The twisting generates
a nontrivial Scherk-Schwarz potential on the moduli space and for
certain twists is equivalent to introducing internal fluxes of various
gauge fields on the torus.

We consider the extension of duality twisting to the full quantum
string theory and discuss the general properties of the resulting
Scherk-Schwarz potential. The global symmetry $G$ of the low energy
effective action is not a symmetry of the quantum theory but is broken
to a discrete U-duality group $G(\Z)$ \refs{\HullYS} that acts on the
integral lattice of $p$-brane charges. Therefore, the twists that can
be lifted to string theory must belong to the duality group $G(\Z)$ 
\mf. This restriction leads to a quantization condition on the mass
parameters appearing in the Scherk-Schwarz potential \mf.

As we review in \S{2.1}, the physically inequivalent twists are
classified by conjugacy classes of $G(\Z)$. We analyze the
Scherk-Schwarz potential and show that the effective low energy
physics of the compactified theory is completely determined by the
conjugacy class, resolving an apparent paradox. Given a potential on
the moduli space, the next question is whether the potential has any
minima and what the structure of the theory is at these minima.

We will see that the task of finding the minima is simplified
considerably by some elegant group theoretic considerations. We
illustrate this point in \S{3} by means of an explicit example in
which the duality twists belong to $SL(2, \Z)$ and outline the
generalization to other groups in \S{4.4}. We show that the minima of
the Scherk-Schwarz potential are in one-to-one correspondence with the
fixed points in the moduli space under the action of the twist
group. One implication of this result is that for a compactification
twisted by an element of the T-duality group, the theory at a minimum
of the potential has an exact conformal field theory description as an
orbifold of a toroidal compactification. The orbifold theory as usual
contains additional twisted sector states that are not visible in the
supergravity analysis. When the twist is {\it not} a perturbative
symmetry, there is no CFT construction for such theories, but the
supergravity analysis and the group theoretic considerations
concerning the minima of the potential can still be applicable.

One motivation for this work is its bearing on the stabilization of
moduli in string theory. The vacuum manifold of string
compactifications is characterized by several moduli that govern the
shape and size of the compactification space as well as the value of
the coupling constant in string theory and correspond to unwanted
massless fields in spacetime. There are stringent observational
constraints on the presence of such massless scalars and even in a
cosmological context the presence of moduli is problematic
\Bank. It is thus interesting to seek string compactifications
with few or no moduli already at the tree level. 

A number of apparently unrelated methods have been utilized in the
literature for constructing models with a small number of moduli.
Compactifying with duality twists or internal fluxes is one way to
stabilize the moduli. In this framework, the twists or the fluxes
generate a nontrivial potential on the moduli space. As a result, the
expectation values of the moduli fields are fixed at the minima of the
potential and many moduli acquire mass 
\refs{\ScSc, \PolchinskiSM  \MichelsonPN \GukovYA \GreeneGH
\deBoerPX \DasguptaSS \CurioSC \GiddingsYU \CurioAE \LouisUY 
\KachruHE \FreyHF \LouisNY  {--} \BeckerSX}.  
This mechanism has been used, for example, to construct models where
all complex structure moduli of Calabi-Yau and torus compactifications
of Type-II and Type-I compactifications are stabilized
\refs{\GiddingsYU, \KachruHE}. Another way to stabilize the moduli is
to orbifold the theory by a symmetry that exists only for special
values of the moduli \MuWi. The moduli are then fixed to take these
special values. In this case, typically there are many additional
massless scalar fields in the twisted sectors. These twisted moduli
can in turn be made massive by including a shift in the orbifolding
action. Using this mechanism for certain special asymmetric orbifolds,
it is possible to construct models where {\it all} moduli except the
dilaton are stabilized \refs{\HMV \DiSi \DabholkarKV {--} \GanorZE}.

In this paper we investigate the relation between these various
approaches. As we will see, in many respects compactifications with
duality twists and internal fluxes are closely related to certain
orbifolds with shifts.

We review and develop the relevant aspects of compactification with
duality twists in \S{2} and illustrate the main points in \S{3} with
an example with $SL(2)$ twists. We discuss the relation between
duality twisting, fluxes, and orbifolds in \S{4} and conclude in \S{5}
with some comments.

\newsec{Compactification with Duality twists}

\subsec{General Formalism}

For simplicity, we consider twisted  reduction on a circle but
these results can be readily extended to   more general toroidal
compactifications.

Consider a $D+1$ dimensional supergravity (or theory of matter coupled
to gravity) with a global symmetry
$G$. An element $g$ of the symmetry group acts on a generic field
$\psi$ as $\psi \to g[\psi]$. Consider now a dimensional reduction of
the theory to $D$ dimensions on a circle of radius $R$ with a periodic
coordinate $y\sim y+ 2 \pi R$. In the twisted reduction, the fields are not
independent of the internal coordinate but are chosen to have a
specific dependence on the circle coordinate $y$ through the ansatz
\eqn\ansatz{
\psi(x^{\mu}, y) = g(y) \, [ \psi (x^{\mu})] }
for some $y$-dependent group element $g(y)$. An important restriction on
$g(y)$ is that the   reduced theory in $D$ dimensions should
be independent of $y$. This is achieved by choosing 
\eqn\element{ g(y)= \exp \left( {My \over 2\pi R} \right) }
for some Lie-algebra element $M$. The map $g(y)$ is not periodic
around the circle, but has a {\it monodromy}
\eqn\mono{ {\cal M} (g)= \exp M.}
The Lie algebra element $M$ generates a one-dimensional subgroup $L$
of $G$. 

It has been seen in explicit examples that Scherk-Schwarz reduction of
a supergravity gives rise to a gauged supergravity; see e.g.
\refs{\gage, \KaMy, \cow, \HullWG, \bergnew}. 
It is easy to see that this must always be the case. Consider a field
$\psi$ in the $D+1$ dimensional theory that transforms in some
representation of $G$ as $\delta \psi = \epsilon
\overline M \psi$ where $\epsilon$ is an infinitesimal parameter and
$\overline M$ is the matrix representation of the element $M$. It is
straightforward to show that on twisted dimensional reduction to $D$
dimensions, the derivative of $\psi$ is replaced by the gauge
covariant derivative $\nabla\psi = d \psi +A \overline M
\psi$, where $A$ is the 1-form gauge potential arising from the
Kaluza-Klein reduction of the metric on the circle. This follows from
demanding general coordinate invariance under transformations of the
form $y \to y + \delta y(x)$ where $x$ are the coordinates of the
noncompact D-dimensional spacetime. We thus obtain a gauged
supergravity where $L$ has become a local symmetry whose gauge field
is the Kaluza-Klein vector potential.  The gauged supergravity has
fermion mass terms and modifications of the fermion supersymmetry
transformations which are linear in the mass matrix $M$, and a scalar
potential (discussed below) which is quadratic in $M$. If any other
vector fields in the theory are singlets under $G$, then the gauge
group is the one-dimensional group $L$ (or strictly speaking the
product of $L$ with the gauge group for the other vector fields, which
is abelian in most of the examples of interest here).  However, if
there are $n$ other abelian gauge fields $A^b$ ($a,b=1,...,n$)
transforming in some representation of $G$, $\delta A^a= \epsilon
\overline M^a{}_b  A^b$, then the gauge group is the semi-direct product
of $L$ with $U(1)^n$ 
with generators $t_y,t_a$ 
and structure constants $f_{ya}{}^b = -f_{ay}{}^b=\overline M^b{}_a$,  
where $t_y$ is the generator corresponding to the Kaluza-Klein vector
field and all other structure constants vanish.

The Scherk-Schwarz ansatz \ansatz\ breaks the global symmetry $G$ down
to the subgroup that commutes with $g(y)$.  Acting with a general
constant element $h$ in $G$ will change the twist to $h g(y) h^{-1}$
and would seem to give a new theory. However, this theory is related to the
original one via the field redefinition $\psi \to h [\psi]$ for all
fields $\psi$, so that the two choices of $g(y)$ in the same conjugacy
class give equivalent reductions related by field-redefinitions \refs{
\mf}.

The map $g(y)$ is a local section of a
principal fiber bundle over the circle with fiber $G$ and monodromy
${\cal M} (g)$ in $G$.  Such a bundle is constructed from $I\times G$,
where $I$ is the interval $[0, 2\pi R]$, by gluing the ends of the
interval together with a twist of the fibers by the monodromy ${\cal
M} $. Two such bundles with monodromy in the same $G$-conjugacy class
are equivalent. 

In classical supergravity, any twist in $G$ is allowed, but in
M-theory, the twists must belong to the duality group $G(\Z)$ and thus
the inequivalent twisted reductions will be classified by the
conjugacy classes of the discrete group $G(\Z)$ \mf.  Monodromies in
$G(\Z)$ related by $G$ conjugation define theories with equivalent
actions, but in general the action of $G$ changes the charge
lattice. For a fixed charge lattice, the equivalent classes of
theories are defined by the classes of $G(\Z)$ monodromy related by
$G(\Z)$ conjugation \mf.

Note that in performing twisted reductions, it is not necessary that the
potential have any critical points, or that the  theory
have a solution which is flat space or (anti-) de Sitter space in $D$
dimensions. For example, in the twisted reduction of IIB supergravity in
\refs{ \Berg} the resulting $D=9$ theory has a potential without
critical points and so has no Minkowski or maximally symmetric vacua.
However, it does have half-supersymmetric domain wall solutions, which
can be lifted to solutions of the 10-dimensional IIB theory, as can
any other solution of the $D=9$ theory. This is a typical situation,
and it is useful to discuss reduction in generality without specifying
a $D$-dimensional solution.

Going around the circle many times generates twists that are powers of
the monodromy $\CM$. We will refer to the discrete abelian subgroup of $G(\Z)$
generated by the monodromy $\CM$ as the twist group of the bundle. If
the order of the twist group is a finite integer $n$, then the
$n$-fold cover of this fiber bundle is trivial because all twists can
be completely undone around a larger circle. That is, with the ansatz
\ansatz\ and \element\ and twist group $\Z _n$, if the range of
$y$ is extended to run from $0$ to $2\pi n R$, then the $n$-fold cover
of the original circle is the circle with the identification $y
\sim y+2\pi n R$ and the monodromy for this covering circle is the
identity, as $\CM ^n=\II$.

As we explain in \S{2.3}, the low energy effective action of the
gauged supergravity in $D$ dimensions is completely determined by the
mass matrix $M$ for a given monodromy $\CM$.  This leads to an
apparent paradox. It is clear from Eq. \mono\ that a given monodromy
matrix can arise in general from infinitely many different mass
matrices $M$ \mf. As the bundle space is determined completely by the
monodromy, different choices of $M$ with the same $\CM$ should give
equivalent theories.  On the other hand, as the mass matrix $M$
appears explicitly in the gauged supergravity action, different
choices of $M$ would appear to give different theories.  For example,
in the case of trivial reduction with $\CM=\II$, there are infinitely
many mass matrices $M$ satisfying $e^M=\II$, each of which would give
a different supergravity action.  We describe in the next subsection
how this ambiguity is resolved.

\subsec{An Ambiguity}

Consider the example of a complex scalar field $\phi$ reduced on a
circle with coordinate $y$ with the identification $y \sim y+2\pi R$.
For a trivial reduction, one has the mode expansion
\eqn\mode{\phi (x,y) = \sum _n e^{iny/R} \phi _n(x),}
giving an infinite set of fields 
$ \phi _n(x)$ in the reduced theory
with mass $m_n \propto n/R$, so that $\phi_0$ is a massless field
and the other modes are massive Kaluza-Klein modes.
If the original theory is invariant under $U(1)$ phase rotations 
$\phi \to e^{i\alpha}\phi$, 
one can include a $U(1)$ twist in the reduction, so that
the $1\times 1$ mass matrix is $M= i m/R$ for some real number $m$, with
monodromy
$\CM= e^{2\pi im}$.
Then
the twisted mode sum becomes
\eqn\mode{\phi (x,y) = \sum _n e^{i(n+m)y/R} \tilde \phi _n(x),}
so that the new modes $ \tilde \phi _n(x)$ have mass
$\tilde m_n \propto (n+m)/R$.
Clearly, if $m$ is an integer, then the two mode sums are equivalent, with
$\tilde \phi _n= \phi _{n+m}$, and the full Kaluza-Klein spectra are the
same, as one would expect from the fact that both reductions have  
monodromy matrix $\CM=\II$.
However, in the twisted case the mass matrix is non-trivial.
This means that if one reduces and then
truncates to the
$n=0$ sector, one is left with a single scalar field
$\tilde \phi _0(x)$
with mass $m/R$, with different masses for different choices of integer
$m$.
In this way, one could  truncate the Kaluza-Klein spectrum to
any one of the massive modes
$\phi_m=\tilde \phi _0$ instead of the usual choice $\phi_0$.
 Similarly, two non-integral choices of mass $m=m_1,m=m_2$
which differ by an integer would give equivalent Kaluza-Klein spectra, but
if one truncated to the $n=0$ sector, one would obtain distinct
truncations.

This applies more generally.
The twisted compactifications are classified by the 
monodromy matrices,
up to conjugation.
Different choices of mass matrix which give equivalent monodromies
will give equivalent Kaluza-Klein spectra, but
can give distinct truncations to the 
\lq zero-mode' sector (the analogue of the $n=0$ sector in the example above
whose only dependence on the extra coordinates comes from the twist).
These different truncations will give different potentials
as they depend on the mass matrix explicitly.
However, in deriving low-energy effective physics, it is important to
choose the
truncation to the lightest fields.
In the   example above, the tower of Kaluza-Klein fields
 $ \tilde \phi _n(x)$ have mass
$\tilde m_n \propto (n+m)/R$
and one could truncate to a single scalar for any given value of
$n$. However, the lightest scalar is for that value of $n$ which minimizes
$|m+n|$ and in deriving the effective low-energy physics, it is important
to choose that value of $n$ if one truncates, so that the effective theory
describes the lightest states.

\subsec{The Scalar Potential}

The moduli fields, which we generically denote by $\Phi$, are not
massless in the reduced theory in general and   there is a nontrivial
Scherk-Schwarz potential $V(\Phi)$ on the moduli space. It is
straightforward to extend the analysis of Scherk and Schwarz \ScSc\
and later generalizations to obtain an explicit formula for this
scalar potential in terms of the mass matrix $M$.  For the case in
which the scalars in $D+1$ dimensions take values in a coset $G/K$
(typically $G$ is a non-compact group with a maximal compact group
$K$) they can be represented by a vielbein $\CV (x)\in G$ transforming
under rigid $G$ transformations and local $K$ transformations as $\CV
\to k(x) \CV g $.
Here we will restrict ourselves to the case in which $\CV $ is a real
matrix in a real representation
of $G$; the  generalization to complex representations is straightforward. 
  The kinetic term is
\eqn\kin{ L= -{1\over 2} {\rm Tr} [\CV^{-1} D_m \CV \CV^{-1} D^m \CV ]}
where $D_m$ is a $K$-covariant derivative with $K$-connection given in
terms of $\CV$ and its derivative.  In this formulation, the theory
has a rigid $G$ symmetry and a local $K$ symmetry. The local $K$
symmetry can be fixed to remove the unphysical degrees of freedom in
$\CV$.  Let $\eta $ be a constant $K$-invariant metric (for
semi-simple $K$, it can be taken to be the Cartan-Killing metric, and
for the standard case in which $K$ is compact, a Lie algebra basis can
be chosen so that $\eta = \unit$). Then one can define the
$K$-invariant field $\h = \CV^t \eta \CV$ transforming under $G$ as
$\h \to g^t \h g $, so that the kinetic term becomes
\eqn\kina{ L= + {1\over 2} {\rm Tr} [\partial_m \h^{-1} \partial^m \h ].}
It is straightforward to show that dimensional reduction on a circle
with a twist determined by the mass matrix $M$ yields a potential in
$D$ dimensions given by
\eqn\pot{ V(\Phi) = e^{a \phi}{\rm Tr} [ M^2 + M^t \h(\Phi) M
\h^{-1}(\Phi)],}
where $e^\phi$ is the modulus corresponding to the radius of the
circle and $a=6/(D-1)(D-2)$. 
The potential arises from the $y$-derivatives in Eq. \kina\ 
with the Scherk-Schwarz ansatz $\CH(\Phi(x), y) = \CM^t(y)
\CH(\Phi(x)) \CM(y)$
with $\CM(y) = \exp{ {M y \over 2\pi R}}$.
The matrix $M$ has dimensions of mass and
introduces mass parameters into the theory. This generalizes the
results of \refs{\ScSc,\ort,\cow}.

One immediate question is whether this potential has any stable minima
and which moduli acquire mass at these minima.  In terms of $\tilde M
= \CV M \CV^{-1}$, the potential becomes
\eqn\pottwo{ V(\Phi) = e^{a \phi}{\rm Tr} [  \tilde M^2 +\tilde  M^t \eta
\tilde  M
\eta^{-1}].}
For a given mass matrix $M$, the potential depends on the moduli
$\Phi$ that parametrize the coset through the matrix ${\tilde
M}(\Phi)$. The dependence on $\phi$ is only through the exponential
factor, so the potential will be stationary with respect to variations
of $\phi$ only if $ V(\Phi) = 0$, which requires either $a
\phi=-\infty$, or $\tilde M =\tilde M _0$ with
\eqn\sfsd{{\rm Tr} [ \tilde M_0 ( \tilde M_0 +  \eta ^{-1}
\tilde  M_0^t
\eta )]=0.}

Let us now restrict to  the   case in which $K$ is compact
and $\eta$ is the  identity matrix (e.g.
$G = SL(N)$ and $K = SO(N)$).  Then the potential can be rewritten as
\eqn\potthree{ V(\Phi) ={1\over 2} e^{a \phi}{\rm Tr} (Y^2)}
where $Y$ is the real symmetric matrix, $Y \equiv [\tilde M + \tilde
M^t]$.  The potential is then manifestly positive, $V(\Phi) \ge 0$
because $Y$ is diagonalizable with real eigenvalues, so that ${\rm Tr}
(Y^2)$ is the sum of the squares of the eigenvalues.  It is clear that
the potential will vanish at a point $\Phi = \Phi_0$ in the moduli
space if and only if $Y$ vanishes at that point.  At such a point
$\Phi_0$ at which $Y=0$, $\tilde M (\Phi_0)$ equals a rotation
generator $\tilde M_0$ with $\tilde M_0 = -\tilde M_0^t$. Moreover,
from the positivity of the potential, the point $\Phi_0$ is a global
minimum that is stable or at least marginally stable. Given such an
antisymmetric $\tilde M_0$, the relation $\tilde M _0 = \CV _0 M
\CV^{-1}_0$ determines the corresponding value $\CV_0$ of the
vielbein $\CV$ at the point $\Phi =\Phi_0$.  To summarize, the only
critical points of the potential for finite $\phi$ are the stable
minima where the potential vanishes and where $\tilde M (\Phi_0)$
is a rotation generator.

We now derive some general properties of the critical points of this
potential which will play a vital role in understanding the relation
between twisted reductions and orbifolds. We will show that the
critical points (or submanifolds) are fixed under the action of the
twist group. The relevant mathematics will be discussed further in
\S{4.4}. Consider then the case in which the mass matrix is
$G$-conjugate to a rotation generator $r$, $r=- r^t$, so that
\eqn\sfsaad{
M=S^{-1}r S
}
for some constant $S\in G$.  Then the monodromy $\CM= e^M$ is
conjugate to a rotation matrix $R = e^r$ satisfying $R^t R =
\unit$,
\eqn\rott{
\CM=S^{-1}R S.
}
The potential now will have a global minimum at the point $\Phi_0$ in
moduli space such that $\CV (\Phi_0) = S$ because at that point $\ti
M_0= r$ and so $Y(\Phi_0) = 0$. At this point, the coset metric takes the value
$\CH_0=S^tS$. This is invariant under the action of the twist group, $\CH_0 \to
\CH'_0 \equiv \CM^t \CH_0 \CM = \CH_0$, as is easily seen using \rott\
and $R^t R = \unit$. Thus, such a critical point is a fixed point
under the action of the twist group generated by $\CM$.

There is a natural action of $G$ on the theory, inherited from the
structure of the $D+1$ dimensional theory, but it is not a symmetry in
$D$ dimensions, as the mass terms and potential are not invariant
under $G$ (although they are preserved by a subgroup). Acting with $G$
is a field redefinition, and there are two situations to
consider. First, if the $D+1$ dimensional theory is a field theory
with a global $G$ symmetry (e.g. a classical supergravity), then the
field redefinition from acting with $G$ takes the $D$-dimensional
theory to an equivalent theory, written in terms of different
variables. The second case is that in which the $D+1$ dimensional
theory has only a $G(\Z)$ symmetry (as in string theory or M-theory
compactifications, or in a classical Kaluza-Klein reduction on 
$\bT^n$ where the massive Kaluza-Klein modes break the
low-energy $SL(n,\R)$ to $SL(n,\Z)$).  If there is a charge lattice
acted on by $G$ and preserved by the subgroup $G(\Z)$, then for a
fixed charge lattice, only field redefinitions from the action of
$G(\Z)$ will lead to equivalent theories.

Since $G$ acts transitively on the coset, any point on the coset
$\Phi_0$ can be moved to any other point $\Phi_0'$ by right
multiplication of the vielbein by some element $U \in G$, $\CV(\Phi_0)
\to\CV(\Phi_0')= \CV(\Phi_0) U$. Under this action, the twist $\CM$
will go to $ \CM' = U^{-1} \CM U$, changing the potential to a new
one.  If $\Phi_0$ was a critical point of the original potential, then
$\Phi_0'$ is a critical point of the new one.  In the first situation
in which $G$ is a symmetry in $D+1$ dimensions, this action of $G$ is
a field redefinition and leads to an equivalent theory and by
acting with $G$, any given critical point $\Phi_0$ can be moved to any
desired point in moduli space $\Phi_0'$. In the second situation in
which the original theory only has a $G(\Z)$ symmetry, acting with $G$
in general takes the theory to an inequivalent one, but acting with
$G(\Z)$ leads to an equivalent theory.  Thus acting with $G$ can move
a critical point to any desired point in moduli space, but in general
changes the theory.  Acting with $G(\Z)$ will take the theory to a
physically equivalent one, and change the monodromy to another
representative of the same $G(\Z)$ conjugacy class.  The $G(\Z)$
action can be used to move any critical point to one in a fundamental
domain $G(\Z)\backslash G/K$ of the moduli space. However, then acting
with $G$ to move it to another point in the same fundamental domain
would lead to an inequivalent theory.

The distinction between these two situations will be important later
when we discuss orbifolds in \S{4}. Different points in the moduli
space where different orbifold theories are possible can be moved to
each other by $G$ transformations and would appear to be equivalent in
the naive low-energy analysis unless we correctly incorporate the
integrality of charges as above by allowing only $G(\Z)$
transformations.

\newsec{Examples with $SL(2)$ Twists}

We now illustrate the main ingredients of this construction by means
of an example of a standard reduction on $\bT^2 $ followed by a
twisted reduction on $\bS^1$. Reducing first on the $\bT^2$ gives a
theory whose symmetries include the mapping class group $SL(2, \Z)$ of
the torus.  One can then reduce further on the circle with a twist
that belongs to this $SL(2, \Z)$. This example will also prepare the
background for establishing the connection with orbifolds, and is
closely related to the IIB compactifications considered in
\refs{\mf, \ort, \cow, \mee, \HullWG, \bergnew}.

\subsec{Pure Gravity}

Consider first a theory of pure gravity with Einstein-Hilbert action
in $D+3$ dimensions. Dimensionally reducing on $\bT^2$ gives a theory
in $D+1$ dimensions whose massless spectrum contains  the graviton, two
Kaluza-Klein gauge bosons and three scalar fields coming from the   moduli of
the torus. The area of the torus $e^\psi$ parametrizes $\R^+$ and the complex
structure $\tau$ of the torus parametrizes $SL(2, \Z) \backslash
SL(2,\R)/SO(2)$.  The $SL(2,\Z)$ is the  group of large
diffeomorphisms of the torus and is a discrete gauge symmetry.  

The truncated massless theory in $D+1$ dimensions now has $SL(2,\R)$
global symmetry and we can consider the reduction on a further circle
to $D$ dimensions with an $SL(2,\R)$ twist. There are three distinct
twisted reductions corresponding to the three distinct $SL(2,\R)$
conjugacy classes
\refs{\mf}.  These are the hyperbolic, elliptic and parabolic
$SL(2,\R)$ conjugacy classes, represented by the monodromy matrices
\eqn\monc{
{\cal M}_h=\pmatrix{ e^m & 0 \cr 0 & e^{-m} \cr }, \qq {\cal
M}_e=\pmatrix{
\cos m & \sin m \cr - \sin m & \cos m \cr } , \qq
{\cal M}_p=\pmatrix{ 1 & m \cr 0 & 1 \cr }
}
respectively,  generated by the matrices
\eqn\gens{
{ M}_h=\pmatrix{ m & 0 \cr 0 & -m \cr }, \qq { M}_e=\pmatrix{ 0 & 
m \cr  -  
m & 0 \cr } , \qq { M}_p=\pmatrix{ 0 & m \cr 0 & 0 \cr }
}
and each class is specified by a single coupling constant or mass
parameter $m$.

For each of these theories the Scherk-Schwarz potential
\pot\ takes a simple form. The scalars 
$\psi,\tau = \tau_1 + i \tau_2$ take values in $GL(2,\R)/SO(2)$ and can
be represented by the $GL(2,\R)$ matrix $\CV$ with a local $SO(2)$
invariance removing one of the four degrees of freedom of $\CV$. Then
$\h= \CV^t \CV$ can be given in terms of $\psi,\tau$ as $\h=e^{\psi
}H(\tau) $ where
\eqn\metric{ H(\tau) \equiv {1\over \tau_2} 
\pmatrix{ 1 & \tau_1 \cr 
\tau_1 & |\tau|^2 \cr }}
 and the potential  is given
by
\eqn\newpot{ V(\tau) = e^{a \phi } {\rm Tr} [   M^2 + {M^t}{H}(\tau) 
{M}
H^{-1} (\tau) ].}
Note that the potential is independent of $\psi$.  For the elliptic
twisting with monodromy $\CM_e$, the potential has a minimum at $\tau
=i$ giving a Minkowski vacuum.  For the parabolic case, the potential
is proportional to $m^2 e^{a \phi + b \Phi}$ where $\tau _2=
e^{-\Phi}$ and and $b$ is a constant, and so the only critical points
are when $a \phi + b \Phi=- \infty$.  For finite $\phi$, this
corresponds to $\tau =i\infty$, representing a degenerate torus.  The
hyperbolic case has no critical points on the upper half plane.

The $SL(2, \R)$ global symmetry of the massless reduction is broken 
down to an $SL(2,\Z)$ subgroup if
the massive Kaluza-Klein states are kept.  For the reduction of the full
Kaluza-Klein theory including the massive states, therefore, the
monodromy must belong to $SL(2,\Z)$.  The $SL(2,\Z)$ conjugacy classes
have been analyzed in \refs{\DeWolfeEU, \DeWolfePR}. For any conjugacy
class $\cM$, $-\cM $ and $\pm \cM^{-1}$ also represent conjugacy
classes, so for each $\cM$ in the following list, there are also
conjugacy classes $-\cM $ and $\pm \cM^{-1}$.

Apart from the trivial class $\cM=\II$, there are four conjugacy classes
that generate twist groups of finite order
\eqn\mond{ {\cal M}_{2}=\pmatrix{ -1& 0 \cr 0 & -1 \cr }, \quad 
{\cal M}_{3}=\pmatrix{0 & 1 \cr -1 & -1 \cr }, \quad
{\cal M}_{4}=\pmatrix{ 0 & 1 \cr - 1 & 0 \cr }, \quad
{\cal M}_{6}=\pmatrix{ 1 & 1 \cr -1 & 0 \cr }.}
The matrices ${\cal M}_{2}, {\cal M}_{3}, {\cal M}_{4}, {\cal M}_{6}$
respectively generate ${\Z_2}, {\Z_3}, {\Z_4}, {\Z_6}$ subgroups
of $SL(2,\Z)$ and the subscript gives the order of
the subgroup. The monodromies ${\cal M}_{3}, {\cal M}_{4}, {\cal
M}_{6}$ are all in the elliptic conjugacy class of $SL(2,\R)$ with
$\vert Tr({\cal M})\vert <2$.

The monodromies in the parabolic and hyperbolic conjugacy classes all
generate twist groups of infinite order. There are an infinite number
of parabolic $SL(2,\Z) $ conjugacy classes with $Tr({\cal M})=2$,
represented by $T^n$:
\eqn\mondp{
{\cal M}_{T_n}=\pmatrix{ 1 & n \cr 0 & 1 \cr }}
with a distinct conjugacy class for each integer $n$. 

There are   an infinite number of hyperbolic $SL(2,\Z)
$ conjugacy classes with $\vert Tr({\cal M})\vert >2$, represented by
\eqn\mondh{{\cal M}_{H_n}=\pmatrix{ n & 1 \cr -1 & 0 \cr }, 
 }
for integers $n$ with  $|n| \ge 3$, together with
sporadic monodromies ${\cal M}(t)$ of trace $t$
\eqn\monds{\eqalign{
{\cal M}({8})&=\pmatrix{ 1 & 2 \cr 3 & 7 \cr },  \qq {\cal 
M}({10})=\pmatrix{ 1
&  4 \cr 2 & 9 \cr } , \qq  {\cal M}({12})=\pmatrix{ 1 & 2 \cr 5 & 11 \cr 
}
\cr 
{\cal M}({13})&=\pmatrix{ 2 & 3 \cr 7 & 11 \cr 
},\qq {\cal M}({14})=\pmatrix{ 1 & 2 \cr 6 & 13 \cr 
}, \dots
\cr}}
and this gives the complete list of sporadic classes for
$3\le t\le 15$.

The  mass matrices corresponding to the monodromies \mond\ and \mondp\ 
are given by
\eqn\mass{\eqalign{
{M}_{2}={\pi} A ^{-1}\pmatrix{ 0 & 1 \cr -1 & 0 \cr } A,\qquad 
{M}_{3}=& {2\pi \over 3\sqrt{3} }\pmatrix{ 1 & 2 \cr -2 & -1 \cr }, \qquad 
{M}_{4}= {\pi \over 2}\pmatrix{ 0 & 1 \cr - 1 & 0 \cr }, \cr
{M}_{6}= {\pi \over 3 \sqrt{3} }\pmatrix{ 1 & 2 \cr -2 & -1 \cr }, \qquad  
{M}_{T^n}=&\pmatrix{ 0 & n \cr 0 & 0 \cr } .
\cr}}
where $A$ is an arbitrary $SL(2, \R)$ matrix. 

The ambiguity discussed in section \S{2.2} arises here from the
infinitely many solutions of the equation $e^M =\II$ given by $M =
{2\pi }\pmatrix{ 0 & n \cr - n & 0 \cr }$. This ambiguity does not
affect the full physical spectrum and in \mass\ we have chosen, for
each monodromy, a simple representative for the mass matrix from the
infinite number of possible choices. Note that after accounting for
this ambiguity, the mass matrices for the monodromies ${\cal M}_{3},
{\cal M}_{4}, {\cal M}_{6}$ are uniquely determined but there are
still an infinite number of mass matrices $M _2$, characterized by the
arbitrary matrix $A$, that all give rise to the same monodromy
$\CM_2$.  Note that changing $A$ is an $SL(2,R)$ conjugation and so a
field redefinition in the truncated theory in which the Kaluza-Klein
modes are absent and the $D+1$ dimensional theory has an $SL(2,\R)$
symmetry, but for the full theory it changes the theory unless it is
an $SL(2,\Z)$ conjugation.  We shall return to the role of $A$ in our
discussion of orbifolds.  Each of the mass matrices \mass\ is
$SL(2,\R)$-conjugate to the mass matrix $M_e$ in \gens, $M_n =U^{-1}M_eU$
and so the corresponding potentials each have a unique critical point
at which $V=0$, and this is located at the image of $\t =i$ under the
action of the $SL(2,\R)$ transformation $U$.

\subsec{Bosonic String}

Consider next the bosonic string compactified on $\bT^2$. In addition
to the metric, we now also have a dilaton and an antisymmetric tensor
among the massless fields. The global symmetry group is $G = O(2, 2)$
and for fixed value of the dilaton, the moduli space of these
compactifications is given by the Narain coset $O(2, 2; \Z) \backslash
O(2, 2)/O(2) \times O(2)$.

A convenient parametrization of this space is in terms of the complex
structure modulus $\tau$ and the complexified K\"ahler modulus
$\sigma$. The real part of $\sigma$ is the area of the torus and the
imaginary part is the value of the 2-form field $B_{mn}$ on the torus.
The moduli space for complex structures is $SL(2, \Z)\backslash
SL(2,\R)/SO(2)$ as before and the K\"ahler modulus parametrizes an
identical space.  The total moduli space is thus
\eqn\twomod{ [SL(2, \Z) \backslash SL(2, \R)/SO(2) \times SL(2, \Z)
\backslash SL(2,\R)/SO(2) ] / \Z_2.}
 The additional $\Z_2$ comes from the ``parity'' element of $O(2, 2,
\Z)$ with determinant $-1$.
This element changes the sign of one of the left-moving
coordinates of the torus and hence corresponds to T-duality along that
coordinate; it exchanges $\tau$ and $\sigma$ and interchanges the two
$SL(2, \Z)$ factors (see, for example, \GiveonFU).

We can now reduce the theory further on a circle with a duality twist
given by a conjugacy class of $G (\Z)= [(SL(2,
\Z)_{\tau}\times SL(2, \Z)_{\sigma}] \rsemidir \Z_2$. 
The subscripts are added to denote that $SL(2, \Z)_{\tau}$ and
$SL(2,\Z)_{\sigma}$ act on $\tau$ and $\sigma$ respectively. The
twists that belong to the $SL(2, \Z)_{\tau}$ factor have already been
discussed in the previous subsection; there are distinct theories
corresponding to each of the conjugacy classes of $SL(2,\Z)$. The
twists by $SL(2, \Z)_{\sigma}$ are nongeometric but are conjugate by
the $\Z_2$ T-duality element to $SL(2, \Z)_{\tau}$ and lead to
equivalent theories. Twisting simultaneously by elements of the two
$SL(2)$ factors with a mass matrix 
\eqn\mplus{
M=(M_\ss \otimes \unit) \oplus (\unit
\otimes M_\tau)
}
 where $M_\ss$ and $M_\tau$ are mass matrices of
$SL(2)_{\sigma}$ and $SL(2)_{\tau}$ twists respectively, results in new
theories. As we discuss in \S{4.1}, these new theories are related to
asymmetric orbifolds.

\subsec{Supergravity}

For a supergravity with a global symmetry $G$ and local symmetry $K$,
with scalars in $G/K$ parametrized by $\CV$, the fermions are inert
under $G$ but transform under $K$.  In a physical gauge in which the
$K$ symmetry is fixed, a $G$ transformation is accompanied by a
compensating $K$ transformation which acts on the fermions.  Given the
low energy action for the massless bosons, the effective action for
the fermions is determined by supersymmetry. Corresponding to the
nontrivial scalar potential \pot, the fermions acquire
moduli-dependent mass terms that are linear in the mass matrix $M$,
and the supersymmetry transformations of the fermions are modified by
terms linear in $M$.

Consider the Scherk-Schwarz reduction from $D+1$ to $D$ dimensions on
a circle, in the formalism in which the local $K$ symmetry is not
fixed.  For the bosonic sector, the reduction is specified by the
choice of a twist in $G$.  In the fermionic sector, there is a choice
of spin structure for the fermions on the circle (i.e. the possibility
of including a twist by $(-1)^F$). The fermions can be decomposed into
$K$ representations, and in principle it is possible to choose a
different spin structure for each $K$ representation.  In addition,
there is the possibility of accompanying this by a twist in $K$.

Alternatively, one can first choose a physical gauge eliminating the
local $K$ symmetry, and then reduce with a twist in $G$ (which acts on
fermions through the compensating transformation) and a choice of spin
structure for each $K$ representation. In the cases that we have
discussed so far, the symmetries include a rigid $SL(2,\R)\subseteq G$
symmetry and a local $U(1)\subseteq K$ in $D+1$ dimensions. In this
case, if we fix the $K$ symmetry completely by choosing physical
gauge, the $SL(2,\R)$ transformation represented by the matrix
\eqn\fgfd{\lll=
\pmatrix{ a & b \cr c & d \cr }
}
will act on a fermion $\ll$ of $U(1)$ charge $q$ by the compensating
$U(1)$ transformation
\eqn\fermions
{\ll \to
\left( {c \bar \tau +d
\over
c  \tau +d}
\right) ^{q/4} \ll.}

Here we restrict ourselves to the case in which we twist only by the
global group $G$  and the spin structure is periodic for all fermions. This
gives reductions specified by a mass matrix $M$ which reduce to the standard
reduction when $M=0$.

In the standard reduction on $\bT^2 $ followed by a twisted reduction
on $\bS^1$ that we have considered above, all gravitini become massive
at the minima of the scalar potential and the supersymmetry is
completely broken.  This can be checked directly, and  will become
apparent once we make the connection with orbifolds.  In the orbifold
description, the gravitini have nontrivial transformations under the
twist groups and are thus projected out, so that there are no massless
gravitini in the spectrum
and   supersymmetry is completely broken.  It is
straightforward, however, to construct models with supersymmetric minima by
compactifying on higher dimensional tori; we will discuss a simple example in
section \S{4.2}.

\subsec{Superstrings}

For the heterotic string on $\bT^2 $, there are additional gauge
fields and extra moduli from the Wilson lines. The Narain moduli space
is now $O(2, 18; \Z)\backslash O(2, 18)/O(2) \times O(18)$.  On the
submanifold of this moduli space where all Wilson lines are turned
off, the duality symmetry is again $[(SL(2, \Z)_{\tau}\times SL(2,
\Z)_{\sigma}] \rsemidir \Z_2$. In this special case, the analysis is
similar to that for the bosonic string. More general reductions
twisted by conjugacy classes of the full duality group $O(2, 18; \Z)$
are quite interesting and are related to heterotic compactifications
with various magnetic fluxes turned on, as will be discussed elsewhere.

For the Type-IIA superstring on $\bT^2 $, the U-duality group is
$SL(3,\Z) \times SL(2,\Z)$.  The $SL(3)$ is a symmetry of the
supergravity action and contains
$SL(2)_\tau$, while the $SL(2)$
factor is only a symmetry of the supergravity equations of motion and
is the $SL(2)_\sigma$ factor considered above. The perturbative
T-duality symmetry is $[(SL(2, \Z)_{\tau}\times SL(2,
\Z)_{\sigma}]$. Note that the $\Z_2$ element corresponding to
T-duality along one leg of the torus is no longer a symmetry because
it interchanges Type-IIA with Type-IIB.  The Type-IIB superstring
compactified on $\bT^2 $ gives the same $D=8$ theory, but now for   IIB it is
$SL(2)_\sigma$ that is contained in $SL(3)$, while the $SL(2)$ factor that is
only a symmetry of the equations of motion is the geometric symmetry
$SL(2)_\tau$.  Whereas in the heterotic or bosonic case, twisting by
$SL(2)_\tau$ or $SL(2)_\sigma$ gave equivalent theories related by T-duality,
in the type II case they give rise to two distinct $SL(2)$ twistings. In the
first, the IIA theory is twisted by $SL(2)_\tau$, and this is T-dual to
twisting Type-IIB by $SL(2)_\sigma$. This results in a theory similar to the
bosonic and the heterotic cases.  In the second, the IIA theory is
twisted by $SL(2)_\sigma$, and this is T-dual to twisting Type-IIB by
$SL(2)_\tau$.  In this case, the twist is by a symmetry that acts via
duality and is only a symmetry of the equations of motion, not of the
action. This results in some novel features, which will be analyzed in
\refs{\Oz}.

For Type-II strings there are other more general possibilities when
the twisting is nonperturbative and the monodromy is an arbitrary element of
$SL(3,\Z) \times SL(2,\Z)$.
For example, the type IIB string in $D=10$ has a nonperturbative
$SL(2,\Z)_\lambda$ symmetry that acts on the dilaton-axion field $\lambda$.
After reducing on $\bT^2$ this $SL(2,\Z)_\lambda$ becomes a subgroup of
$SL(3,\Z)$ and is conjugate to the perturbative $SL(2,
\Z)_{\sigma}$ discussed above. Therefore, the $SL(2,\Z)_\lambda$
twists are dual to the $SL(2,\Z)_\sigma$ twists. Even though the group
theoretic considerations are identical in the two cases, the
realization in terms of perturbative string modes will be quite
different. For example, twists that correspond to turning on NS-NS
fluxes will be conjugate to twists that correspond to turning on R-R
fluxes.

Note that the $D=7$ theory obtained by twisting with an element of the
$SL(2,\Z)_\lambda$ can also be obtained by first reducing the IIB
theory on a circle with an $SL(2,\Z)_\lambda$ twist $\CM$ to $D=9$,
and then performing a standard reduction on $\bT^2$.  Thus, the $D=7$
theories obtained by twisting with $SL(2,\Z)_\lambda$ are precisely
the $\bT^2$ reductions of the $D=9$ theories of
\refs{\ort, \cow,\mf,\HullWG,\bergnew} and have a very similar structure.  
The $D=9$ theory can be thought of as F-theory compactified on a
$\bT^2$ bundle over $\bS^1$ with monodromy $\CM$ \refs{\mf}.

\newsec{Orbifolds, Duality twists, and Fluxes}

Given a theory with a discrete symmetry $\bX$, its orbifold is
obtained by gauging the symmetry. The Hilbert space of the orbifold
consists of states of the original theory that are invariant under
$\bX$, together with new twisted string states that are closed up to a
nontrivial $\bX$ transformation.  We will be interested in orbifolds
of strings compactified on $\bT^2 \times \bS^1$. For special values of
the torus modulus, the torus will be invariant under a discrete $\Z_n$
symmetry of finite order $n=2,3,4$ or $6$. For such a torus, the
orbifold group $\bX=\Z_n$ relevant for our purpose is generated by a
$\Z_n$ generator of the torus symmetry group accompanied by an order
$n$ shift along the circle.

\subsec{Bosonic String}

Let us first consider orbifolds of the bosonic string where the
discrete rotation is geometric and acts symmetrically on the
left-moving and right-moving coordinates of the torus.  To see what
geometric rotations are allowed, let $z$ be the complex coordinate of
$\bT^2$ with the identifications $z \sim z+1
\sim z + \tau$, where $\tau$ is the complex structure modulus of the
torus. For what follows, the K\"ahler modulus can be arbitrary so the
over-all scale of the torus is not important. Associated with the
torus is a lattice of points in the complex plane, $\{ z=m + n \tau
\}$, for arbitrary integers $m$ and $n$. Now, a rotation in the
complex plane becomes a symmetry of the torus only if it is a symmetry
of the lattice. A $\Z_2$ rotation through $\pi$ that takes $z$ to $-z$
is a symmetry of all lattices. Additional symmetries are possible for
special lattices (i.e. for special values of $\tau$) given by the
crystallographic classification \DixonJW.  A square lattice with $\tau
= i$ has an enhanced $\Z_4$ symmetry generated by the rotations $z
\rightarrow e^{i\pi/2} z$ and a hexagonal lattice with $\tau = e^{2\pi
i/3}$ has an enhanced $\Z_6$ symmetry generated by $z \rightarrow
e^{i\pi/3} z$ with a $\Z_3$ subgroup generated by $z \rightarrow
e^{2i\pi/3} z$.  The only possible discrete rotation symmetries of the
torus are $\Z_2, \Z_3, \Z_4, \Z_6$.

The orbifold action for our purposes will be one of these $\Z_n$
rotations of a torus at a special value of the modulus with a
simultaneous order $n$ shift along the circle of radius $nR$ for $n=2,
3, 4, 6$.  Note that the list of allowed orbifold rotations is in one--to--one
correspondence with the list of twist groups generated by the monodromies
$\CM_2, \CM_3, \CM_4, \CM_6$ that we encountered earlier in a rather different
context. We now explain the relation between the orbifolds and the twisted
reductions.

It is   clear that all of the above orbifolds can be viewed
as twisted reductions. The group $SL(2, \Z)$ of large diffeomorphisms
of $\bT^2$ has a natural action on the lattice defining the torus and
the $\Z_n$ symmetry of a special lattice is   a subgroup of $SL(2,
\Z)$ that leaves the lattice invariant. Conjugation by
$SL(2, \Z)$ gives a physically equivalent rotation and thus again
there is a dependence only on conjugacy classes.  If the circle has
radius $r$ and coordinate $y \sim y+2\pi r$, then the orbifolded
theory is identified under the action of a $\Z_n$ rotation accompanied
by a shift $y \to y+2\pi r/n$.  This is equivalent to the twisted
reduction on a circle of radius $R=r/n$ with a twist by the $\Z_n$
generator.  Since the orbifold satisfies the string equations of
motion with vanishing ground state energy at tree level, the
Scherk-Schwarz potential must have a stable (or marginally stable)
minimum with zero energy at this point.

The converse is more interesting and less obvious. Compactification
with a duality twist is more general than the orbifold construction in
certain respects because it can be carried out without restricting the
moduli to special values and the moduli can have nontrivial variation
along the circle and in the $D$-dimensional spacetime. Moreover, we
can twist by any monodromy, giving distinct theories for each of the
infinite number of conjugacy classes listed in \S{3}.  The orbifold,
on the other hand, is possible only for special values of the moduli
where the lattice admits a symmetry and the class of allowed orbifold
rotations is finite. As we now discuss, the connection between the two
is provided by the Scherk-Schwarz potential. The minima of the
potential occur precisely at the fixed points in the coset space
$SL(2)/SO(2)$ under the action of the twist group, and these are
precisely the points in moduli space where orbifolding is possible.

Consider first the parabolic and the hyperbolic conjugacy classes of
$SL(2, \Z)$. Monodromies in these conjugacy classes generate twist
groups of infinite order and have no fixed points on the upper half
plane with $\tau_2$ strictly positive and finite. As discussed in
\S{3}, the Scherk-Schwarz potential has no stable minima with $\tau_2$
strictly positive and finite in these cases, consistent with the fact that
there is no standard orbifold formulation in this situation.

Monodromies in the elliptic conjugacy classes of $SL(2,\Z)$ generate
twists of finite order. As they are $SL(2,\R)$-conjugate to a
rotation, they must have a fixed point.  In fact, it follows from a
theorem given in \S{4.4} that any finite order subgroup of $G(\Z)$
always has a fixed point on $G/K$ for any non-compact semi-simple $G$
with $K$ its maximal compact subgroup. Moreover, together with the
discussion in \S{2.3} this implies that the Scherk-Schwarz potential
for a given elliptic monodromy has a stable minimum precisely at this
fixed point. We now check these facts by hand for the simple case of
$SL(2)$ by explicitly finding the minima of the potential for the mass
matrices given by \mass.

When $G= SL(2)$, the vielbein can always be written in the physical
gauge as an upper triangular matrix with the parametrization
\eqn\vielebein{
\CV(\tau) = {1\over \sqrt{\tau_2}} \, \pmatrix{1 & \tau_1 \cr 0 & \tau_2}}
so that the metric $\CH = \CV^t \CV$ takes the canonical form
\metric. (That this can always be done is seen most easily by using
the Iwasawa decomposition of a general $SL(2)$ matrix as a
product $k \CV$ where $k$ is an $SO(2)$ matrix and $\CV$ is an upper
triangular matrix and then fixing the physical gauge to gauge
away $k$.)  In this parametrization, given an arbitrary mass matrix 
$M = \pmatrix{ -d & b \cr c & d}$ in the Lie algebra of $SL(2,\R)$, the matrix
$\tilde M = \CV M \CV^{-1}$ is given by
\eqn\masstilde{
\tilde M =   {1\over \tau_2} 
\pmatrix{1 & \tau_1 \cr 0 & \tau_2}  
\pmatrix{-d & b \cr c & d}
\pmatrix{\tau_2  & -\tau_1 \cr 0 & 1}
=  {1\over \tau_2}
\pmatrix{ -d\ttwo + c\tone\ttwo & d\t_1 + b -c\tone^2 + d\tone\cr
c\ttwo^2 & -c\tone\ttwo + d\ttwo}}

Now we have seen in \S{2.3}, the potential can be written in the form
\eqn\potfour { V(\tau) ={1\over 2} e^{a \phi}{\rm Tr} (Y^2)}
where $Y$ is a real symmetric matrix, $Y \equiv [\tilde M + \tilde M^t]$.
Therefore, for a given mass matrix $M$, a minimum occurs precisely
for those values of $\tau$ for which the corresponding $Y$ matrix vanishes.
The $Y$ matrices corresponding to the four mass matrices in \mass\ 
for the elliptic conjugacy classes are given by
\eqn\ymatrix{\eqalign{
Y_2 = & \pmatrix{ -d\ttwo + c\tone\ttwo & 
d\t_1 + b  - c \tone^2 + c\ttwo^2 + d\tone \cr
d\t_1 + b  - c \tone^2 + c\ttwo^2 + d\tone &
-c\tone\ttwo + d\ttwo} \cr
Y_3  = & {4 \pi \over 3 \sqrt{3} \ttwo} 
\pmatrix{ \ttwo -2 \tone\ttwo & 1  + \tone^2 - \ttwo^2  -\tone \cr
 1 + \tone^2 - \ttwo^2  -\tone  & - \ttwo + 2\tone\ttwo} = 2 Y_6  \cr 
Y_4  = & {\pi \over 2 \ttwo} 
\pmatrix{ -2\tone\ttwo &  1  + \tone^2 - \ttwo^2  \cr
 1  + \tone^2 - \ttwo^2 & 2\tone\ttwo} \cr 
}}
Note that in the matrix $Y_2$, the three real numbers $b, c, d$ are
subject to the constraint $d^2 + bc = -\pi^2$ and thus it depends
effectively on only two parameters. This follows from the fact that the
mass matrix $M_2$ in \mass\ depends on an arbitrary $SL(2, \R)$ matrix
$A$ and is an arbitrary trace-less matrix whose determinant equals
$\pi^2$.

Now, the minima of the potential can be readily found.  The matrices
$Y_3$ and $Y_6$ vanish only at $\t =\exp{(\pi i/3)}$ and thus for
twists by the monodromies $\CM_3$ and $\CM_6$, the minimum of the
potential \potfour\ occurs precisely at points where a $\Z_3$ and
$\Z_6$ orbifold action is possible. Similarly the matrix $Y_4$
vanishes only at $\tau =i$ and thus for the monodromy $\CM_4$ the
potential has a minimum precisely where a $\Z_4$ orbifold is possible.

For the conjugacy class $\CM_2$, 
the position at which the matrix $Y_2$ vanishes
depends on the choice of the
numbers $b, c, d$ in \ymatrix,
corresponding to the choice of the $SL(2, \R)$ matrix
$A$ in \mass.
Choosing $A=1$, $d=0, b=-c$, $Y_2$ vanishes at $\tau=i$.
Now conjugating with $U\in SL(2,\R)$ 
gives $A=U$ and can be used to move the point at which $Y_2$ vanishes 
to any desired point in moduli space.
Changing $A$ in this way changes the compactified theory unless $A\in
SL(2,\Z)$, and this $SL(2,\Z)$ redundancy can be used to move the critical
point into a fundamental domain.
This freedom is
consistent with the fact that a $\Z_2$ orbifold is possible for all values
of $\t$ and is a consequence of the fact that the orbifold twist in
this case belongs to the center of the duality group.

We can understand the existence and the location of these minima more
succinctly following the discussion \S{2.3} in a way that will be
generalized to other twist groups in \S{4.4}. Every monodromy $\CM_n$
of finite order $n$ has $\vert Tr(\CM_n)\vert <2$ and   is in the
elliptic $SL(2, \R)$ conjugacy class $\CM_n$, so that it is conjugate
to the rotation matrix $\CM_e$ given in (3.1), for some value of the
angle of rotation $m$.  Moreover, since $(\CM_n )^n =\unit$, the angle
must be $m={2\pi N\over n}$ for some integer $N$.  The monodromies in
(3.5) are in fact conjugate to the rotation matrix $R_n$, where $R_n$
is the  $SO(2)$ rotation through ${2\pi \over n}$, i.e. there
exists a (constant) $SL(2, \R)$ matrix $S_n$ such that
\eqn\rot{
S_n \CM_n S_n^{-1} = R_n}
Note that given an $S_n$ that solves this equation,
left-multiplication by an arbitrary $SO(2)$ matrix $k$ gives another
matrix $S_n'\equiv k S_n$ that also solves this equation. We can use
this gauge freedom to bring all matrices $S_n$ to an upper triangular
form. For the cases $n=2,3,4,6$, the matrices $S_n$ are given by
\eqn\conjugacy{
S_2 = V, \quad S_4 = \pmatrix{ 1 & 0 \cr 0 & 1}, \quad S_3 = S_6 
= \sqrt{{2\over \sqrt{3}}} \pmatrix{ 1 & \half  \cr 0 & {\sqrt{3} \over 2}}.}
Note that $S_2$ is an arbitrary $SL(2, \R)$ upper triangular matrix
$V$ because $M_2$ depends on an arbitrary $SL(2,
\R)$ matrix $A$ which can written  as a product $A = k V$ where $k$ is
an $SO(2)$ matrix.

For these monodromies, the mass matrix $M_n$ can be chosen (using the
ambiguity discussed in \S{2.2}) so that after this conjugation it becomes   the
rotation generator %
\eqn\genrot{
S_n M_n S_n^{-1} = {2\pi \over n} \pmatrix{ 0 & -1 \cr 1 & 0}.}
We have seen in \S{2.3} that for such a mass matrix, the
Scherk-Schwarz potential has a global minimum at $\CV=S_n$ at which
the potential vanishes, and that this is a fixed point under the
action of the twist group generated by $\CM_n$.

We thus conclude that for an elliptic duality twist $\CM_n \in SL(2,
\bZ)_{\tau}$, the critical points of the Scherk-Schwarz potential
are precisely at the fixed points of the twist. The potential vanishes
at the minimum and the theory at the minimum is a symmetric orbifold
of the type discussed above using the twist group generated by $\CM_n$
accompanied by a shift.

Orbifolds with twisted boundary conditions around toroidal
directions have been considered before, for example, in 
\refs{\RohmAQ, \MuWi, \FerraraES, \FerraraJX}, usually with boundary 
conditions that break supersymmetry. Our analysis illuminates the
place of such orbifold conformal field theories in the string
configuration space. If we climb up the Scherk-Schwarz potential from
the minimum, the string equations of motion will no longer be
satisfied and there would be no CFT description of the theory because
we have perturbed the CFT by an irrelevant perturbation. Nevertheless,
from the spacetime point of view, it is a mild way of going off-shell
with operators that correspond to massive fields in spacetime with
masses of order of the inverse radius of the circle and our analysis
gives the off-shell potential.

Duality twists that belong to $SL(2, \Z)_{\sigma}$ are related to the
one above by a T-duality along one of the legs of the torus. The most
general case when we twist by an arbitrary element of $O(2, 2; \Z)$
would therefore twist the coordinate and the T-dual coordinate
independently of each other. The minima of the potential in this case
would be described by the most general order $n$ asymmetric orbifold
with an asymmetric rotation of the torus accompanied by a shift along
the circle.

The possible asymmetric rotations can be easily classified
\DabholkarKV\ and are given by the automorphisms of the Lorentzian 
lattice $\Gamma^{2,2}$ for special values of the moduli that are left
fixed by the twists.  There are fixed planes for the cases that we
have already discussed when the T-duality twist acts only on $\tau$ or
only on $\sigma$.  There are also fixed points in the general case
that have more symmetry.  For example, the point $\sigma=i, \tau=i$
has an enhanced $(\Z_4 \times \Z_4) \rsemidir \Z_2$ symmetry, the point
$\sigma=\tau=\rho$ with $\rho=e^{
\pi i/3}$ has an enhanced $\Z_9$ symmetry and the point $\sigma=i$,
$\tau=\rho$ (or vice versa) has a $\Z_{12}$ symmetry which acts
quasicrystallographically \HMV\ on the lattice. At any of these points
in the moduli space, a $\Z_n$ subgroup of the symmetry can be combined
with an order $n$ shift to obtain an asymmetric orbifold. This
orbifold would describe the theory at the minimum of the potential in
the corresponding Scherk-Schwarz reduction, with mass matrix of the
form \mplus.

\subsec{Superstrings}

In the case of superstrings, the action of the orbifold rotation must be
lifted to spacetime fermions. Consider, for example, a $\bT^2$
reduction along the $X^8$ and $X^9$ directions. The torus coordinate
$z$ can be written as $X^8 + iX^9$, and the $\Z_n$ rotations discussed
in the previous section are generated by elements $\exp{({2\pi i
J_{89}/ n})}$ where $J_{89}$ is the generator of rotations in the $89$
plane.  When spacetime fermions are present, the eigenvalues of
$J_{89}$ are half-integral and $\exp {(2\pi i J_{89})} = (-1)^F$ where
$F$ is the fermion number; as a result these rotations now generate
$\Z_{2n}$ groups of order $2n$. For odd $n$, an order $n$ symmetry
generated by $\exp{({2\pi i J_{89}/n})} (-1)^F$ is also possible.  We
suppose there is a further circular direction $X^7$ say, and orbifold
by these transformations combined with the appropriate shifts in the
$X^7$ coordinate.

These orbifolds break supersymmetry completely because in the light
cone Green-Schwarz formalism (with $X^8,X^9$ both transverse
coordinates), no components of $Spin(8)$ spinors are left invariant by
the rotation in the $89$ plane. When the radius of the $X^7$ circle is
of string scale, all these models contain tachyons in the twisted
sector and are unstable. However, for a large enough circle there will
be no tachyons and the twisted states will be very massive. This is
the regime in which one can compare the orbifolds with the
supergravity analysis of compactifications with duality twists given
in the previous sections.

The above applies to orbifolds based on subgroups of $SL(2,
\Z)_{\tau}$.  For the heterotic string, the ones based on $SL(2,
\Z)_{\sigma}$ are related by T-duality and are very similar.  
For the Type-IIA string,   orbifolds  by subgroups of $SL(2,
\Z)_{\sigma}$ are distinct from  orbifolds  by subgroups of 
$SL(2, \Z)_{\tau}$ and are T-dual to orbifolds of Type-IIB 
by subgroups of $SL(2, \Z)_{\tau}$.

When the duality twist does not belong to the T-duality group then the
theory at the minimum of the Scherk-Schwarz potential cannot be
described by a perturbative orbifold, but the supergravity analysis of
\S{2} and \S{3} is still applicable. For example, in the supergravity
analysis the twists that correspond to turning on Ramond-Ramond fluxes
are on the same footing as those that correspond to turning on NS-NS
fluxes (see below for a discussion of fluxes in this context). The group
theoretic considerations of this and the previous sections can be equally
well applied to such nonperturbative twists, in particular for finding the
minima of the Scherk-Schwarz potential.

For the standard reduction on $\bT^2 $ followed by a twisted reduction
on $\bS^1$ of Type-IIB, all nonperturbative twists belong to
$SL(3)$. If we restrict attention to the nonperturbative $SL(2,
\Z)_{\lambda}$, then the considerations are similar to those for
$SL(2, \Z)_{\sigma}$. The monodromy $\CM_2$ actually corresponds
to a perturbative symmetry $\Omega (-1)^{F_L}$ where $\Omega$ is
orientation reversal and ${F_L}$ is the left-moving fermion number
\DabholkarZD. Therefore, modding out by this symmetry gives rise to a
perturbative orientifold. The orientifold has no orientifold planes or
D-branes because of the shift along the circle. The $\CM_3, \CM_4,
\CM_6$ twists are nonperturbative and the Scherk-Schwarz potential
will fix the dilaton-axion field $\lambda$ to either $i$ or $ e^{\pi
i/3}$ where the string would be strongly coupled. The classical
 analysis given here can still be reliable in such situations in the
spirit of F-theory \refs{\VafaXN},  
especially if the theory at the minimum preserves enough
supersymmetry. Since this $SL(2, \Z)_{\lambda}$ is conjugate to $SL(2,
\Z)_{\sigma}$ by an element of $SL(3)$ we expect that the theories at
the minima with nonperturbative twists will be dual to the
perturbative orbifolds discussed above by using the adiabatic argument
\VafaGM. 

It is easy to construct models with unbroken supersymmetries by
compactifying on higher tori of dimensions $2N$ and choosing a duality
twist that is a subgroup of $SU(N)$. The resulting orbifold theory at
the minimum then has $SU(N)$ holonomy and preserves some number of
supersymmetries. As a simple example that illustrates this point,
consider Type-IIB on a $\bT^4 \times \bS^1$. We take the twists to
be in $SL(4, \Z)$ which is the group of large
diffeomorphisms of $\bT^4$. The simplest nontrivial conjugacy class is
the element $-\unit$ that generates a twist group of order
two. Because it is a twist of finite order, the Scherk-Schwarz
potential will have a stable minimum and the $\Z_2$ symmetry of the
orbifold theory at the minimum is generated by the reflection of all
coordinates of $\bT^4$ accompanied by a half-shift along the
circle. Note that without the half-shift, the $\bT^4 /\Z_2$ orbifold
would have given us a $\bK_3$ and we would have obtained a standard
Type-IIB compactification on $\bK_3 \times \bS^1$ to five dimensions
with sixteen unbroken supersymmetries. When the orbifolding action
includes the half-shift, one would still obtain a theory in five
dimensions with sixteen supersymmetries, but all twisted states will
now be massive. In particular, the vector multiplets that come from
the sixteen fixed points of the reflection on $\bT^4$ will now be
massive thereby stabilizing all moduli that belong to these multiplets
as well as the moduli in the untwisted sector that are projected out
by the orbifolding.

\subsec{Relation to turning on Fluxes}

In this subsection we explain the relation between the twisted
reductions and compactifications with internal fluxes.

The toroidal compactification on $\bT^2$ followed by this twisted
reduction on an $\bS^1$ is equivalent to reducing on a three-manifold
$B$ which is the total space of the torus bundle over a circle with
metric
\eqn\bunmet {ds_B^2=
(2\pi R)^2 dy^2 +{{\cal A} \over \tau_2}\vert dx_1 +\tau (y) dx_2 \vert ^2
}
where the fiber is a $\bT^2$ with real periodic coordinates $x_1,x_2$,
$x_i \sim x_i +1$, constant area modulus ${\cal A}$ and complex
structure $\tau(y)$, which depends on the coordinate $y$. The twisted
reduction on the circle with the ansatz $\tau (y) =\tau _{g(y) }$
associated with a particular torus bundle $B$ is precisely the
compactification on the three dimensional total space $B$ \mf.  For
the parabolic conjugacy class, $\tau(y)= \tau_1+i \tau _2 + n y$ where
$m$ is the integral mass parameter in \mondp, and $\tau_1,\tau_2$ are
independent of $y, x_i$.  Then the metric is
\eqn\bunmettr {ds_B^2=
(2\pi R)^2 dy^2 +{{\cal A} \over \tau_2}
( dx_1 + { A})^2
+{\cal A}\tau_2 dx_2 ^2
}
where
${  A} = (\tau_1 + n y) dx_2$.
The  total space can also be regarded as a
circle bundle over a 2-torus
\mf, with fiber coordinate $x_1$, base space coordinates $y,x_2$ and 
connection 1-form ${ A}$ and first Chern number $n$.  We thus see that
the parabolic conjugacy class $\CM_{T_n}$ corresponds to turning on
$n$ units of magnetic flux of the Kaluza-Klein gauge field.
T-dualizing in the $x_1$ fiber direction untwists the bundle to give a
torus metric on $\bT^3$
\eqn\bunmettr {ds_B^2=
(2\pi R)^2 dy^2 +{\tau_2 \over  {\cal A}}
 dx_1 ^2
+{\cal A}\tau_2 dx_2 ^2
}
but turns on a $B$-field with field strength $H= n dx_1 \wedge dx_2
\wedge dy$ corresponding to a constant $H$-flux over $\bT^3$.

For the elliptic conjugacy classes, the orbifold at the minimum of the
potential can be viewed as turning on magnetic flux tubes similar to
the non-compact Melvin solutions \refs{\RussoCV, \RussoIK,
\TakayanagiJJ}. In the non-compact Melvin solution, the orbifolding action is
a rotation in a plane accompanied by a shift along a circle and this
orbifold can be interpreted as a Melvin background with magnetic flux
of the Kaluza-Klein vector potential. The total flux in the plane is a
function of the angle of rotation in the plane and since the angle is
continuous, the flux can be changed continuously. By contrast, in the
situation that we discuss in this paper, the rotation angle is
quantized because we are rotating the coordinate of a torus and not of
a plane. As we have seen, the only allowed rotation angles for $\bT^2$
are $\pi/3$, $\pi/2$, $\pi$, and $2\pi/3$ and consequently only a
finite number of discrete values of the flux are allowed.

For the hyperbolic cases, the situation is more complicated and it is
unclear whether there is a relation of the reduction to a toroidal
reduction with flux.

\subsec{Generalizations}

Generalizations to higher duality groups are very interesting and can
be used to fix moduli in a more realistic context preserving some
supersymmetry. We will not analyze   explicit models here but
instead present a number of general results that are useful for the
analysis of the Scherk-Schwarz potential in these cases. 

We consider a theory with a moduli space $G(\Z)\backslash G/K$ with
$G$ non-compact semi-simple and $K $ the maximal compact
subgroup.\footnote{$^\dagger$}{Because $K$ acts on the left and
$G(\Z)$ on the right in our conventions in this paper, the coset
should be denoted by $K\backslash G /G(\Z)$; however, with a slight
abuse of notation, we adhere to the common usage,
denoting the moduli space by $G(\Z)\backslash G/K$.}  Our prime
example will be $G=SL(N,\R)$ and $K=SO(N)$.

For $G(\Z)$ (e.g. $SL(N, \Z)$), many more conjugacy classes are
possible and we will not discuss them explicitly here. One general
question of interest for a given conjugacy class is whether the
Scherk-Schwarz potential has a minimum, and if so, where in the moduli
space it lies. The following theorem is useful for addressing this
question.  See, for example, \group\ for a proof.

{\it Theorem}: Every finite order subgroup $H \subset G(\Z) \subset G$
with $G$ non-compact semi-simple is conjugate to a subgroup of the
maximal compact subgroup $K$. Thus, there exists a matrix $S \in G$
such that $ S H S^{-1} = K_1 \subset K$.

The space $G/K$ is defined as a coset with the equivalence relation $g
\sim k g$ for every $g \in G$ and $k \in K$. If we denote the equivalence 
class of $g$ by $[g]$ then the coset is the set $\{ [g]\}$ of all
equivalence classes. The equivalence class of the identity $[\unit]$
corresponds to the entire group $K$. An element $h$ of $G$ acts on the
coset by right multiplication $[g] \to [gh]$. It is clear from the
equivalence $\unit K = K \unit \sim \unit$ that the point $[\unit]$ in
$G/K$ is a fixed point under the action of $K$ by right
multiplication. Therefore, by the theorem above, every finite order
subgroup $H$ also has a fixed point on $G/K$. This property is closely
related to the fact that the spaces $G/K$ have negative
curvature. Indeed, the equivalence class $[S]$ is the desired fixed
point under right-multiplication by $H$ since $SH = K_1S \sim S$. It
is also clear that since $k^t k =
\unit$, the metric $\CH _0=S^t S$ is invariant under
H-transformations: $h^t S^t S h = S^t S$ for all $h
\in H$. Because $H$ leaves the metric invariant, it defines a symmetry
of the corresponding integer lattice in $\R^N$ and can be used for
orbifolding.

These results imply that any twist $\CM$ that generates a finite order
subgroup $H$  is conjugate by an $SL(N, \R)$ matrix $S$ to an $SO(N)$
matrix. By \mono, it will result in a mass matrix that is conjugate by
$S$ to a rotation generator. We have seen in \S{2.3} that in this case
when mass matrix is conjugate to a  rotation generator, $\CV_0=S$ or $\CH_0
=S^tS$ is a stable minimum of the Scherk-Schwarz potential. Using this
physics input we conclude that for the finite order twists $H \subset
SL(N, \Z)$ the matrix $S$ defines a minimum on the coset of the
Scherk-Schwarz potential at which $V=0$.

\newsec{Conclusions}

Even though we have focused here on duality twists in $\bT^2 \times
\bS^1$ compactifications, these methods can be applied equally well to
more general compactifications on higher tori and other manifolds such
as $\bK3$ and Calabi-Yau threefolds that have interesting duality
symmetries. We have seen that there is a close relation between
compactifications with perturbative duality twists and orbifolds. Our
considerations here are useful even for nonperturbative
duality twists and for duality twists that correspond to turning on
internal RR-fluxes. The structure of duality twists for higher groups
is expected to be much richer because many more conjugacy classes are
possible.  For general twists, the Scherk-Schwarz potential can be
quite complicated and explicit extremization is not easy. However, the
group theoretic considerations discussed here provide an efficient way
for finding the minima and the properties of the theory at the
minima. It would be interesting to elucidate further the relation of
duality twisting with compactifications with internal fluxes and to
see if some of the recent models that fix moduli with fluxes can be
analyzed in this framework.

We have seen that in the Type-II circle compactifications considered
here with $SL(2)$ twists, only the elliptic conjugacy classes lead to
stable minima. However, in more general toroidal compactifications
with higher groups, it is likely that other conjugacy classes also
lead to stable minima. For example, the parabolic conjugacy classes
correspond to turning on H-flux. It is known that in orientifolds of
Type-I on $\bT^6$, if additional orientifold charges are present, the
inclusion of 3-form fluxes can lead to gauged supergravities
\refs{\AndrianopoliAQ, \DAuriaTC} that have stable minima 
\refs{\GiddingsYU, \KachruHE}. It would also be interesting to see 
in the more general cases which twists lead to stable minima. In such
more general situations, the twist groups may have fixed sub-manifolds
instead of fixed points in the moduli space where the potential has a
minimum. In such cases, only some of the moduli will be stabilized.

By considering a U-duality twist that has a unique fixed point on the
moduli space, one can construct models with or without supersymmetry
in any dimension that stabilize all moduli except the radius of the
circle used for twisting. In the framework described here we require
an $\bS^1$ factor for twists but in more general situations where the
manifold of compactification has circle fibration, it might be possible
to twist along this fiber in a way analogous to F-theory
\refs{\VafaXN, \KumarZX, \HellermanAX}. If supersymmetry is broken, the
classical analysis would be quantum corrected but we expect that the
existence and the location of the minima which depend on
considerations of symmetry should still be valid.  It would be
interesting to explore further if these different techniques can be
combined to construct realistic models with few or no moduli.

\bigskip

\centerline{\bf Acknowledgments}\nobreak
\medskip

It is a pleasure to thank John Schwarz, Sandip Trivedi, and especially
T. N. Venkataramana for useful discussions. Some of the work was done
during the `M-theory' workshop at the Issac Newton Institute.

\listrefs

\bye

\end